\documentclass[a4paper,twoside]{article}

\usepackage{epsfig}
\usepackage{subfigure}
\usepackage{calc}
\usepackage{amssymb}
\usepackage{amstext}
\usepackage{amsmath}
\usepackage{amsthm}
\usepackage{dsfont}
\usepackage{multicol}
\usepackage{pslatex}
\usepackage{SCITEPRESS}
\usepackage[small]{caption}
\usepackage[square]{natbib}
\usepackage{url}
\usepackage[pdfpagelabels,colorlinks,urlcolor=blue,linkcolor=blue,citecolor=blue]{hyperref}
\hypersetup{pageanchor=false,
    bookmarksnumbered=true
}

\newcommand{\RR}{{{\rm I \kern -0.2em R}}}
\newcommand{\DD}{{{\rm I \kern -0.2em D}}}
\newcommand{\CC}{{{\mbox{\rm \hspace*{0.05ex}
\rule[.18ex]{.18ex}{1.24ex} \kern -.65em C}}}}
\newcommand{\be}{\begin{equation}}
\newcommand{\ee}{\end{equation}}
\newcommand{\ba}{\left [ \begin{array}}
\newcommand{\ea}{\end{array} \right ]}

\newcommand{\rank}{\mathop{\mathrm{rank}}}
\newcommand{\diag}{\mathop{\mathrm{diag}}}

\newtheorem{theorem}{Theorem}

\begin{document}

\title{Descriptor system techniques and software tools}

\author{\authorname{Andreas Varga }
\affiliation{Gilching, Germany
}
\email{varga.andreas@gmail.com}
}

\onecolumn \maketitle \normalsize \vfill
\section*{Abstract}
{The role of the descriptor system representation as basis for reliable numerical computations for system analysis and synthesis, and in particular, for the manipulation of rational matrices, is discussed and available robust numerical software tools are described. }

\section*{Keywords}
{Modelling; Differential-algebraic systems; Rational matrices; Numerical analysis; Software tools} \\

\subsubsection*{AMS subject classifications:} 34A09, 93C, 93B20, 93B40, 93C05, 93D20

\section{Introduction}
A \emph{linear time-invariant} (LTI) continuous-time descriptor
system is described by the equations
\begin{equation}\label{cdss} \begin{array}{rcl}
E \dot x(t) &=& Ax(t) + Bu(t), \\
y(t) &=& Cx(t) + Du(t),\end{array} \end{equation}
where $x(t) \in \mathds{R}^n$ is the state vector, $u(t) \in \mathds{R}^m$ is the input vector, $y(t) \in \mathds{R}^p$ is the output vector, and $A, E \in \mathds{R}^{n\times n}$, $B \in \mathds{R}^{n\times m}$, $C \in \mathds{R}^{p\times n}$, $D \in \mathds{R}^{p\times m}$.
The square matrix $E$ is possibly singular, but we assume that the linear matrix pencil $A-\lambda E$, with $\lambda$ a complex parameter, is regular  (i.e., $\det (A-\lambda E) \not\equiv 0$).  A LTI discrete-time descriptor system has the form
\begin{equation}\label{ddss}  \begin{array}{rcl}
E x(k+1) &=& Ax(k) + Bu(k), \\
y(k) &=& Cx(k) + Du(k).\end{array} \end{equation}
Descriptor system representations of the forms   (\ref{cdss}) and (\ref{ddss}) are the most general descriptions
of LTI systems. Standard LTI state-space systems correspond to the case $E = I_n$. We will alternatively denote the LTI descriptor systems (\ref{cdss}) and (\ref{ddss}) with the quadruples $(A-\lambda E,B,C,D)$ or $(A,B,C,D)$ if $E = I_n$.

Continuous-time descriptor
systems frequently arise when modeling interconnected systems involving linear differential equations and algebraic relations, and are also common
in modeling constrained mechanical systems (e.g., contact
problems). Discrete-time descriptor representations
are encountered in the modeling of some economic processes.
The descriptor system representation is instrumental in devising general computational procedures (even for standard LTI systems), whose intermediary steps involve operations leading to descriptor representations (e.g.,  system inversion or system conjugation in the discrete-time case).

The  input-output behavior of the LTI systems (\ref{cdss}) and (\ref{ddss}) can be described in the form
\begin{equation}\label{iom} \mathbf{y}(\lambda) = G(\lambda) \mathbf{u}(\lambda), \end{equation}
where $\mathbf{u}(\lambda)$ and $\mathbf{y}(\lambda)$ are the transformed input and output vectors, using, in the continuous-time case, the Laplace transform with $\lambda = s$, and, in the discrete-time case, the $Z$-transform with $\lambda = z$, and where
\begin{equation}\label{tfm} G(\lambda) = C(A-\lambda E)^{-1}B + D \end{equation}
is the \emph{transfer function matrix} (TFM) of the system.
The transfer function matrix $G(\lambda)$ is a rational matrix having entries which are rational functions in the complex variable $\lambda$ (i.e., ratios of two polynomials in $\lambda$).  $G(\lambda)$ is \emph{proper} (\emph{strictly proper}) if each entry of $G(\lambda)$ has the degree of its denominator larger than or equal to (larger than) the degree of its numerator. If $E$ is singular, then $G(\lambda)$ could have entries with the degrees of numerators exceeding the degrees of the corresponding denominators, in which case $G(\lambda)$ is \emph{improper}.
We will use the  alternative notation
\begin{equation}\label{tfm-dss} G(\lambda) := \ba{c|c} A-\lambda E & B \\ \hline C & D \ea \, ,\end{equation}
to relate the  TFM $G(\lambda)$ in (\ref{tfm}) to a particular  quadruple $(A-\lambda E,B,C,D)$.

An important application of LTI descriptor systems is to allow the  numerically reliable manipulation of rational matrices (in particular, also of polynomial matrices). This is possible, because for any rational matrix $G(\lambda)$, there exists a quadruple \linebreak[4] $(A-\lambda E,B,C,D)$ with $A-\lambda E$ regular, such that (\ref{tfm}) is fulfilled. Determining the matrices $A$, $E$, $B$, $C$ and $D$ for a given rational matrix $G(\lambda)$ is known as the realization problem and the quadruple $(A-\lambda E,B,C,D)$ is called a descriptor  realization of $G(\lambda)$. The solution of the realization problem is not unique. For example, if $U$ and $V$ are invertible matrices of the same order as $A$, then two realizations $(A-\lambda E,B,C,D)$ and $(\widetilde A-\lambda \widetilde E,\widetilde B,\widetilde C,D)$ related as
\begin{equation}\label{dssim} \widetilde A-\lambda \widetilde E = U(A-\lambda  E)V, \; \; \; \widetilde B = U B, \; \; \; \widetilde C =  CV,  \end{equation}
have the same transfer function matrix. The relations (\ref{dssim}) define a (restricted) \emph{similarity transformation} between the two descriptor system representations. Performing similarity transformations is a basic tool to manipulate descriptor system representations.
An important aspect is the existence of minimal realizations, which are descriptor realizations with the smallest possible state dimension $n$. The characterization of minimal descriptor realizations in terms of relevant systems properties is done in Section \ref{secMR}. To simplify the presentation, we will assume in most of the cases discussed that the employed realizations of rational matrices are minimal.

Complex synthesis approaches of controllers and filters for plants modeled as LTI systems are often described as conceptual computational procedures in terms of input-output representations via TFMs. Since the manipulation of rational matrices is numerically not advisable because of the potential high sensitivity of polynomial based representations, it is generally accepted that the manipulation of rational matrices is best performed via their equivalent descriptor realizations. In what follows, we focus on discussing a selection of descriptor system techniques which are frequently encountered as computational blocks of the synthesis procedures. Whenever possible, we will indicate the best available numerical algorithms, but refrain from discussing computational details, which can be found in the cited references. We conclude with the presentation of a short overview of available software tools for descriptor systems.

\section{Basics of manipulating rational matrices}

In this section, we present the basic manipulations of rational matrices, which represent the building blocks of more involved manipulations.

\subsection*{Basic operations}
We consider some operations involving a single TFM $G(\lambda)$ with the descriptor realization $(A-\lambda E,B,C,D)$. The \emph{transposed} TFM $G^T(\lambda)$ corresponds to the \emph{dual descriptor system} with the realization
\[ G^T(\lambda) = \ba{c|c}A^T-\lambda E^T & C^T\\ \hline \\[-3mm]B^T & D^T \ea \, .\]

If $G(\lambda)$ is invertible, then an inversion free realization of the \emph{inverse} TFM $G^{-1}(\lambda)$ is given by
\[ G^{-1}(\lambda) = \ba{cc|c} A-\lambda E & B & 0\\C & D & I \\ \hline 0 & -I & 0 \ea \, .\]
This realization is not minimal, even if the original realization is minimal.
However, if $D$ is invertible, then an alternative realization of the inverse is
\[ G^{-1}(\lambda) = \ba{c|c} A-BD^{-1}C-\lambda E & -BD^{-1} \\ \hline \\[-3.5mm] D^{-1}C & D^{-1} \ea ,\]
which is minimal if the original realization is minimal. Notice that this operation may generally lead to an improper inverse even for standard state-space realizations $(A,B,C,D)$ with singular $D$.

The \emph{conjugate} (or \emph{adjoint}) TFM $G^\sim(\lambda)$ is defined in the continuous-time case as $G^\sim(s) = G^T(-s)$ and has the realization
\[ G^\sim(s) = \ba{c|c} -A^T-s E^T & C^T \\ \hline \\[-3mm] -B^T & D^T \ea \, , \]
while in the discrete-time case $G^\sim(z) = G^T(z^{-1})$ and has the realization
\[ G^\sim(z) = \ba{cc|c} E^T -z A^T & 0 & -C^T \\
z B^T & I & D^T \\ \hline 0 & I & 0 \ea \, .\]
If $G(z)$ has a standard state-space realization $(A,B,C,D)$ with $A$ invertible, then an alternative realization of $G^\sim(z)$ is
\[ G^\sim(z) = \ba{c|c} A^{-T} -z I &  -A^{-T}C^T \\ \hline \\[-3mm]
B^TA^{-T} &  D^T -B^TA^{-T}C^T \ea \, .\]
This operation may lead to a conjugate system with improper $G^\sim(z)$, for a standard discrete-time state-space realization $(A,B,C,D)$ with singular $A$.

\subsection*{Basic couplings}

Consider now two LTI systems with the rational TFMs $G_1(\lambda)$ and $G_2(\lambda)$, having the descriptor realizations  $(A_1-\lambda E_1,B_1,C_1,D_1)$ and
 $(A_2-\lambda E_2,B_2,C_2,D_2)$, respectively. The product  $G_1(\lambda)G_2(\lambda)$ represents the \emph{series coupling} of the two systems and has the descriptor realization
\[ G_1(\lambda)G_2(\lambda) := \ba{cc|c} A_1-\lambda E_1 & B_1C_2 & B_1D_2 \\
0 & A_2 - \lambda E_2 & B_2 \\ \hline
C_1 & D_1C_2 & D_1D_2 \ea \, .\]
\index{descriptor system!parallel coupling}%
The \emph{parallel coupling} corresponds to the sum $G_1(\lambda)+G_2(\lambda)$ and has the realization
\[ G_1(\lambda)+G_2(\lambda) := {\arraycolsep=.5mm\ba{cc|c} A_1-\lambda E_1 & 0 & B_1 \\
0 & A_2 - \lambda E_2 & B_2 \\ \hline
C_1 & C_2 & D_1+D_2 \ea} \, .\]
The \emph{column concatenation} of the two systems corresponds to building $\left[\begin{smallmatrix} G_1(\lambda)\\G_2(\lambda) \end{smallmatrix}\right]$  and has the  realization
\[ \ba{c} G_1(\lambda)\\G_2(\lambda) \ea = \ba{cc|c} A_1-\lambda E_1 & 0 & B_1 \\
0 & A_2 - \lambda E_2 & B_2 \\ \hline
C_1 & 0 & D_1 \\
0 & C_2 & D_2 \\
\ea \, .\]
The \emph{row concatenation} of the two systems corresponds to building $\big[\, G_1(\lambda) \; G_2(\lambda)\,\big]$ and has the realization
\[ \big[\, G_1(\lambda) \; G_2(\lambda)\,\big] = {\arraycolsep=.5mm\ba{cc|cc} A_1-\lambda E_1 & 0 & B_1 & 0\\
0 & A_2 - \lambda E_2 & 0 & B_2 \\ \hline
C_1 & C_2 & D_1 & D_2
\ea} \, .\]
The \emph{diagonal stacking} of the two systems corresponds to building $\left[\begin{smallmatrix} G_1(\lambda)&0\\0& G_2(\lambda) \end{smallmatrix}\right]$ and has the realization
\[ {\arraycolsep=.5mm\ba{cc} G_1(\lambda)&0\\0& G_2(\lambda) \ea = \ba{cc|cc} A_1-\lambda E_1 & 0 & B_1 & 0\\
0 & A_2 - \lambda E_2 & 0 & B_2 \\ \hline
C_1 & 0 & D_1 & 0 \\
0 & C_2 & 0 & D_2
\ea }\, .\]

\section{Minimal Realization} \label{secMR}
The manipulation of rational matrices via their descriptor representations relies on the fact that for any rational matrix $G(\lambda) \in \mathds{R}(\lambda)^{p\times m}$, there exist $n \geq 0$ and the  real matrices $E, A \in \mathds{R}^{n\times n}$, $B\in \mathds{R}^{n\times m}$, $C\in \mathds{R}^{p\times n}$ and $D\in \mathds{R}^{p\times m}$, with $A-\lambda E$ regular, such that (\ref{tfm}) holds and $n$ has least possible value.
If $G(\lambda)$ is proper, this fact is a well-known result of the realization theory of standard state-space systems for which numerically reliable minimal realization methods exist. Using this result, a simple realization technique allows to obtain a minimal descriptor realization of a (generally improper) rational matrix $G(\lambda)$ by using the additive decomposition
\[ G(\lambda) = G_p(\lambda) + G_{pol}(\lambda) ,\]
where $G_p(\lambda)$ is the proper part of $G(\lambda)$ and $G_{pol}(\lambda)$ is its strict polynomial part (i.e., without constant term). The proper part $G_p(\lambda)$ has  a standard state-space realization  $(A_p,B_p,C_p,D_p)$  and for the strictly proper TFM $\lambda^{-1} G_{pol}(\lambda^{-1})$ we can build another standard state-space realization $(A_{pol},B_{pol},C_{pol},0)$. Then, we obtain
\[ G(\lambda) =  \ba{cc|c} A_p-\lambda I & 0 & B_p \\ 0 & I-\lambda A_{pol} & B_{pol} \\ \hline
C_p & C_{pol} & D_p \ea \, .\]

A minimal descriptor system realization \linebreak[4] $(A-\lambda E,B,C,D)$  is characterized by the following five conditions.  \begin{theorem}[\citep{Verg81}]\label{T-desc-minreal}
A descriptor system realization  $(A-\lambda E,B,C,D)$ of order $n$ is minimal if the following conditions are fulfilled:
\[ \begin{array}{rl} (i) & \rank\ba{cc} A-\lambda E & B \ea = n, \quad \forall \lambda \in \mathds{C} , \\ \\[-3mm]
(ii) & \rank\ba{cc}  E & B \ea = n, \\ \\[-3mm]
(iii) & \rank\ba{c} A-\lambda E \\ C \ea = n, \quad \forall \lambda \in \mathds{C} , \\ \\[-3mm]
(iv) & \rank\ba{c}  E \\ C \ea = n, \\ \\[-3mm]
(v) & A\mathcal{N}(E) \subseteq \mathcal{R}(E) .
\end{array} \]
\end{theorem}
\noindent Here, $\mathcal{N}(E)$ denotes the (right) nullspace of $E$, while $\mathcal{R}(E)$ denotes the range space of $E$.

The conditions $(i)$ and $(ii)$ are known as \emph{finite} and \emph{infinite controllability}, respectively. A system which fulfills both $(i)$ and $(ii)$ is called \emph{controllable}. Similarly, the conditions $(iii)$ and $(iv)$ are known as \emph{finite} and \emph{infinite observability}, respectively. A system which fulfills both $(iii)$ and $(iv)$ is called \emph{observable}. Condition $(v)$ expresses the absence of non-dynamic modes (see their definition in Section~\ref{sec:adv}). A descriptor realization which satisfies only $(i)-(iv)$ is called \emph{irreducible} (also weakly minimal). The numerical computation of minimal realizations is addressed, for example,  in \citep[Section~10.3.1]{Varg17}.

\section{Canonical Forms of Linear Pencils}

The main appeal of descriptor system techniques lies in their ability to address various analysis and synthesis problems of LTI systems in the most general setting, both from theoretical and computational standpoints. The basic mathematical ingredients for addressing analysis and synthesis problems of descriptor systems are two canonical forms of linear matrix pencils: the \emph{Weierstrass canonical form} of a regular pencil and the \emph{Kronecker canonical form} of a singular pencil. For a given a linear pencil $M-\lambda N$ (regular or singular), the corresponding canonical form $\widetilde M - \lambda \widetilde N$ can be obtained using a pencil similarity transformation of the form $\widetilde M - \lambda \widetilde N = U(M-\lambda N)V$, where $U$ and $V$ are suitable invertible matrices.

If the pencil $M-\lambda N$ is regular and $M, N \in \mathds{C}^{n\times n}$, then, there exist invertible matrices $U \in \mathds{C}^{n\times n}$ and $V \in \mathds{C}^{n\times n}$ such that
\be\label{Weierstrass} U(M-\lambda N)V = \ba{cc} J_f-\lambda I & 0 \\ 0 & I-\lambda J_\infty \ea ,\ee
where $J_f$ is in a (complex) Jordan canonical form
\be\label{Jordan} J_f = \diag \left(J_{s_1}(\lambda_1), J_{s_2}(\lambda_2), \ldots, J_{s_k}(\lambda_k) \right) \, ,\ee
with  $J_{s_i}(\lambda_i)$ an elementary $s_i\times s_i$ Jordan block of the form
\[ J_{s_i}(\lambda_i) = \ba{cccc} \lambda_i & 1  \\ & \lambda_i & \ddots \\ & & \ddots & 1 \\ &&&\lambda_i \ea \]
and $J_\infty$ is nilpotent and has the (nilpotent) Jordan form
\be\label{Jordan-null} J_\infty = \diag \big(J_{s_1^\infty}(0), J_{s_2^\infty}(0), \ldots, J_{s_h^\infty}(0) \big) \, .\ee
The Weierstrass canonical form (\ref{Weierstrass}) exhibits the finite and infinite eigenvalues of the pencil $M-\lambda N$. Overall, by including all multiplicities, there are $n_f = \sum_{i=1}^{k}s_i$ \emph{finite eigenvalues} and $n_\infty = \sum_{i=1}^{h}s_i^\infty$ \emph{infinite eigenvalues}. Infinite eigenvalues with $s_i^\infty =1$ are called \emph{simple infinite eigenvalues}. If $M$ and $N$ are real matrices, then there exist real matrices $U$ and $V$ such that the pencil $U(M-\lambda N)V$ is in a \emph{real} \emph{Weierstrass canonical form}, where  the only difference is that $J_f$ is in a real Jordan form. In this form, the elementary real Jordan blocks correspond to pairs of complex conjugate eigenvalues. If $N = I$, then all eigenvalues are finite and $J_f$ in the Weierstrass form is simply the (real) Jordan form of $M$. The transformation matrices can be chosen such that $U = V^{-1}$.

If $M-\lambda N$ is an arbitrary (singular) pencil with $M, N \in \mathds{C}^{m\times n}$, then, there exist invertible matrices $U \in \mathds{C}^{m\times m}$ and $V \in \mathds{C}^{n\times n}$ such that
\be\label{Kronecker} U(M-\lambda N)V = {\arraycolsep=1mm\ba{ccc} K_r(\lambda) \\ & K_{reg}(\lambda) \\&&K_l(\lambda)\ea} ,\ee
where:
\begin{enumerate}
\item[1)] The full row rank pencil $K_r(\lambda)$ has the form
\[ K_r(\lambda) = \diag \big(L_{\epsilon_1}(\lambda), L_{\epsilon_2}(\lambda), \cdots, L_{\epsilon_{\nu_r}}(\lambda) \big) \, , \]
with  $L_{i}(\lambda)$  ($i \geq 0$) an $i\times (i+1)$  bidiagonal pencil of form
\be\label{Liblocks} L_i(\lambda) = \ba{cccc} -\lambda & 1 \\ & \ddots & \ddots \\ && -\lambda & 1 \ea  \, ; \ee
\item[2)] The regular pencil $K_{reg}(\lambda)$ is in a Weierstrass canonical form
\[ K_{reg}(\lambda) = \ba{cc} \widetilde J_f-\lambda I  \\ & I-\lambda \widetilde J_\infty \ea \, ,\]
with $\widetilde J_f$ in a (complex) Jordan canonical form as in (\ref{Jordan}) and
with $\widetilde J_\infty$ in a nilpotent Jordan form as in (\ref{Jordan-null});
\item[3)] The full column rank $K_l(\lambda)$ has the form
\[ K_l(\lambda) = \diag \big(L^T_{\eta_1}(\lambda), L^T_{\eta_2}(\lambda), \cdots, L^T_{\eta_{\nu_l}}(\lambda) \big) \, .\]
\end{enumerate}

As it is apparent from (\ref{Kronecker}), the Kronecker canonical form  exhibits the right and left singular structures of the pencil $M-\lambda N$ via the full row rank block $K_r(\lambda)$ and full column rank block $K_l(\lambda)$, respectively, and the eigenvalue structure via the regular pencil $K_{reg}(\lambda)$. The full row rank pencil $K_r(\lambda)$ is $n_r\times (n_r+\nu_r)$, where $n_r = \sum_{i=1}^{\nu_r} \epsilon_i$, the full column rank pencil $K_l(\lambda)$ is $(n_l+\nu_l)\times n_l$, where $n_l = \sum_{j=1}^{\nu_l} \eta_j$, while the regular pencil $K_{reg}(\lambda)$ is $n_{reg}\times n_{reg}$, with $n_{reg} = \tilde n_f+\tilde n_\infty$, where $\tilde n_f$ is the number of eigenvalues of $J_f$ and $\tilde n_\infty$ is the number of infinite eigenvalues of $I-\lambda \widetilde J_\infty$ (or equivalently the number of null eigenvalues of $\widetilde J_\infty$).
The normal rank $r$ of the pencil $M-\lambda N$ results as
\[  r := \rank (M-\lambda N) = n_r+\tilde n_f+\tilde n_\infty + n_l .\]
If $M-\lambda N$ is regular, then there are no left- and right-Kronecker structures and the Kronecker canonical form is simply the Weierstrass canonical form.

The reduction of matrix pencils to the Weierstrass or Kronecker canonical forms generally involves the use of non-orthogonal, possibly ill-conditioned, transformation matrices. Therefore, the computation of these forms must be avoided when devising  numerically reliable algorithms for descriptor systems. Alternative condensed forms, as the (ordered) generalized real Schur form of a regular pencil or various Kronecker-like forms of singular pencils, can be determined by using exclusively perfectly conditioned orthogonal transformations and can be always used instead of the Weierstrass or Kronecker canonical forms, respectively, in addressing the computational issues of descriptor systems. Numerically stable algorithms to determine Kronecker-like forms are described in \citep{Varg17} and in the literature cited therein.

\section{Advanced Descriptor Techniques} \label{sec:adv}

In this section we discuss a selection of problems involving rational matrices, whose solutions  involve the use of advanced descriptor system manipulation techniques. These techniques are instrumental in addressing controller and filter synthesis problems  in the most general setting by using numerically reliable algorithms for the reduction of linear matrix pencils to appropriate condensed forms.



\subsection*{Normal rank}
The \emph{normal rank} of a $p\times m$ rational matrix $G(\lambda)$, which we denote by $\rank G(\lambda)$, is the maximal number of linearly independent rows (or columns) over the field of rational functions $\mathds{R}(\lambda)$.
\index{transfer function matrix (TFM)!normal rank}%
It can be shown that the normal rank of $G(\lambda)$ is the maximally possible rank of the complex matrix $G(\lambda)$ for all values of $\lambda \in \mathds{C}$ such that $G(\lambda)$ has finite norm.
 For the calculation of the normal rank $r$ of $G(\lambda)$ in terms of its  descriptor realization $(A-\lambda E,B,C,D)$, we use the relation
\[ r = \rank S(\lambda)-n ,\]
where $\rank S(\lambda)$ is the normal rank of the system matrix pencil $S(\lambda)$ defined as
\be\label{syspencil} S(\lambda) := \ba{cc} A-\lambda E & B \\ C & D \ea  \ee
and $n$ is the order of the descriptor state-space realization.
The normal rank $r$ can be easily determined from the Kronecker form of the pencil $S(\lambda)$ as
\[  r := n_r+n_{reg} + n_l - n, \]
where $n_r$, $n_{reg}$ and $n_l$ defines the normal ranks of $K_r(\lambda)$, $K_{reg}(\lambda)$, and $K_l(\lambda)$, respectively, in the Kronecker form (\ref{Kronecker}) of $S(\lambda)$.

For numerical computations, the Kronecker-like form of the system matrix pencil provides the same structural information by using pencil reduction algorithms based on orthogonal transformations. An even more efficient way to determine the normal rank is to determine the maximum of the rank of $S(\lambda)$ for a few random values of the frequency variable $\lambda$ by using singular values based rank evaluations.

\subsection*{Poles and zeros}
The poles of $G(\lambda)$ are related to $\Lambda(A-\lambda E)$, the eigenvalues of the \emph{pole pencil}  $A-\lambda E$ (also known as the generalized eigenvalues of the pair $(A,E)$).
For a minimal  realization $(A-\lambda E,B,C,D)$ of $G(\lambda)$, the finite poles of $G(\lambda)$ are the $n_{p,f}$ finite eigenvalues in the Weierstrass canonical form of the regular pencil $A-\lambda E$, while the number of infinite poles is given by $n_{p,\infty} = \sum_{i=1}^{h}(s_i^\infty-1)$, where $s_i^\infty$ is the multiplicity of the $i$-th infinite eigenvalue. The infinite eigenvalues of multiplicity ones are the so-called \emph{non-dynamic modes}. The McMillan degree of $G(\lambda)$, denoted by $\delta\big(G(\lambda)\big)$, is the total number of poles $n_p := n_{p,f} + n_{p,\infty}$ of $G(\lambda)$
\[  \delta\big(G(\lambda)\big) := n_p \]
and satisfies $\delta\big(G(\lambda)\big) \leq n$. A \emph{proper}  $G(\lambda)$ has only finite poles.

A proper $G(\lambda)$  is \emph{stable} if all its poles belong to the appropriate stable region $\mathds{C}_s \subset \mathds{C}$, where $\mathds{C}_s$ is the open left half plane of $\mathds{C}$, for a continuous-time system, and the interior of the unit circle centered in the origin, for a discrete-time system. $G(\lambda)$ is \emph{unstable} if it has at least one  pole (finite or infinite) outside of the stability domain $\mathds{C}_s$.

The zeros of $G(\lambda)$  are those complex values of $\lambda$ (including also infinity), where  the rank of the system matrix pencil (\ref{syspencil})
drops below its normal rank $n+r$. Therefore, the zeros can be defined on the basis of the eigenvalues of the regular part $K_{reg}(\lambda)$ of the Kronecker form (\ref{Kronecker}) of $S(\lambda)$. The  finite zeros of $G(\lambda)$ are the $n_{z,f}$ finite eigenvalues of the regular pencil $K_{reg}(\lambda)$, while $n_{z,\infty}$, the total number of  infinite zeros,  is the sum of multiplicities of infinite zeros, which  are defined by the multiplicities of infinite eigenvalues of $K_{reg}(\lambda)$  minus one. The total number of zeros is $n_z := n_{z,f}+n_{z,\infty}$. A proper and stable $G(z)$ is \emph{minimum-phase} if all its zeros are finite and stable.

The number of poles and zeros of $G(\lambda)$ satisfy the relation
\[ n_p = n_z + n_l + n_r , \]
where $n_r$ and $n_l$ are the normal ranks of $K_r(\lambda)$ and $K_l(\lambda)$, respectively, in the Kronecker form (\ref{Kronecker}) of $S(\lambda)$.

Numerically stable algorithms for the computation of poles employ orthogonal transformations to reduce the pole pencil $A-\lambda E$ to a quasi-upper triangular form (i.e., with the pair $(A,E)$ in a generalized Schur form), while for the computation of zeros use orthogonal transformations to reduce the system matrix pencil $S(\lambda)$ to special Kronecker-like forms \citep{Misr94}.

\subsection*{Rational nullspace bases}

Let $G(\lambda)$ be a $p\times m$ rational matrix of normal rank $r$ and let $(A-\lambda E,B,C,D)$ be a minimal descriptor realization of $G(\lambda)$. The set of $1\times p$ rational (row) vectors $\{ v(\lambda) \}$ satisfying $v(\lambda)G(\lambda) = 0$ is a linear space, called the \emph{left nullspace} of $G(\lambda)$, and has dimension $p-r$. Analogously, the set
of $m\times 1$ rational (column) vectors $\{ w(\lambda) \}$ satisfying $G(\lambda)w(\lambda) = 0$ is a linear space, called the \emph{right nullspace} of $G(\lambda)$, and has dimension $m-r$.

The $p-r$ rows of a $(p-r)\times p$ rational matrix $N_l(\lambda)$ satisfying $N_l(\lambda)G(\lambda) = 0$ is a basis of the left nullspace of $G(\lambda)$, provided $N_l(\lambda)$ has full row rank. Analogously, the $m-r$ columns of a $m\times (m-r)$ rational matrix $N_r(\lambda)$ satisfying $G(\lambda)N_r(\lambda) = 0$ is a basis of the right nullspace of $G(\lambda)$, provided $N_r(\lambda)$ has full column rank. The determination of a rational  left nullspace basis $N_l(\lambda)$ of
$G(\lambda)$ can be easily turned into the problem of determining a rational basis of the system matrix $S(\lambda)$. Let $M_l(\lambda)$ be a suitable rational matrix such that
\be\label{yl1} Y_l(\lambda) := [\, M_l(\lambda)\; N_l(\lambda)\,] \ee
 is a left
nullspace basis of the associated system matrix pencil $S(\lambda)$ (\ref{syspencil}).
Thus, to determine $N_l(\lambda)$  we can determine first $Y_l(\lambda)$,  a
left nullspace basis of $S(\lambda)$, and then
$N_l(\lambda)$ results as
\[ N_l(\lambda) = Y_l(\lambda)\ba{c} 0 \\ I_{p} \ea . \]
By duality, if $Y_r(\lambda)$ is a right nullspace basis of $S(\lambda)$, then a right nullspace basis of $G(\lambda)$ is given by
\[ N_r(\lambda) = [ \; 0 \;\; I_m \;] Y_r(\lambda). \]

The Kronecker canonical form (\ref{Kronecker}) of the system pencil $S(\lambda)$ in (\ref{syspencil}) allows to easily determine left and right nullspace bases of $G(\lambda)$. Let $\overline S(\lambda) = US(\lambda)V$ be the Kronecker canonical form (\ref{Kronecker}) of $S(\lambda)$, where $U$ and $V$ are the respective left and right transformation matrices. If $\overline Y_l(\lambda)$ is a left nullspace basis of $\overline S(\lambda)$, then
\be\label{polminbasisleft}  N_l(\lambda) = \overline Y_l(\lambda) U\ba{c} 0 \\ I_{p} \ea . \ee
Similarly, if
$\overline Y_r(\lambda)$ is a right nullspace basis of $\overline S(\lambda)$ then
\be\label{polminbasisright}  N_r(\lambda) = [ \; 0 \;\; I_m \;]  V\overline Y_r(\lambda) . \ee

We choose $\overline Y_l(\lambda)$ of the form
\be\label{polminbasisleft1}  \overline Y_l(\lambda) = \big[ \; 0 \;\; 0 \;\; \overline Y_{l,3}(\lambda) \;\big] ,\ee
where $\overline Y_{l,3}(\lambda)$ satisfies $\overline Y_{l,3}(\lambda)K_l(\lambda) = 0$. Similarly, we choose $\overline Y_r(\lambda)$ of the form
\be\label{polminbasisright1} \overline Y_r(\lambda) = \ba{c} \overline Y_{r,1}(\lambda)\\ 0 \\ 0\ea ,\ee
where $\overline Y_{r,1}(\lambda)$ satisfies $K_r(\lambda)\overline Y_{r,1}(\lambda) = 0$.
Both $\overline Y_{l,3}(\lambda)$ and $\overline Y_{r,1}(\lambda)$ can be determined as polynomial or rational matrices and the resulting bases are polynomial or rational as well.

Numerically reliable computational approaches to compute proper nullspace bases of rational matrices rely on using Kronecker-like forms (instead of the Kronecker form), which can be determined by using exclusively orthogonal similarity transformations. Moreover, these methods are able to determine nullspace bases of minimal McMillan degree and with arbitrary assigned poles \citep{Varg08a}.

\subsection*{Additive decompositions}
Let $G(\lambda)$ be a rational TFM  with a descriptor system realization $G(\lambda) = (A-\lambda E,B,C,D)$. Consider a disjunct partition of the complex plane $\mathds{C}$ as
\be\label{Cgoodbadds}  \mathds{C} = \mathds{C}_g \cup \mathds{C}_b, \quad \mathds{C}_g \cap \mathds{C}_b = \emptyset \, ,\ee
where both $\mathds{C}_g$ and $\mathds{C}_b$ are symmetrically located with respect to the real axis, and $\mathds{C}_g$ has at least one point on the real axis. Since $\mathds{C}_g$ and $\mathds{C}_b$ are disjoint, each pole of $G(\lambda)$ lies either in $\mathds{C}_g$ or in $\mathds{C}_b$.
Using a similarity transformation of the form (\ref{dssim}), we can determine an equivalent representation of
$G(\lambda)$ with partitioned system matrices of the form
\be\label{desc-specdec} \begin{array}{lcl} G(\lambda) &=&  \ba{c|c} UAV-\lambda UEV & UB \\ \hline \\[-3mm] CV & D \ea \\
&=& {\arraycolsep=.5mm\ba{cc|c}  A_g -\lambda E_g & 0 &  B_{g} \\ 0 &  A_b -\lambda E_b &  B_{b} \\ \hline \\[-4mm]  C_g &  C_b & D \ea} \, , \end{array}\ee
where $\Lambda(A_g -\lambda E_g) \subset \mathds{C}_g$ and $\Lambda(A_b -\lambda E_b) \subset \mathds{C}_b$.
It follows that $G(\lambda)$
can be additively decomposed as
\be\label{dss_add}  G(\lambda) = G_g(\lambda) + G_b(\lambda) ,\ee
where
\[ G_g(\lambda) = \ba{c|c}  A_g -\lambda E_g & B_{g} \\ \hline \\[-4mm]  C_g &  D \ea, \] \[ G_b(\lambda) = \ba{c|c}  A_b -\lambda E_b & B_{b} \\ \hline \\[-4mm]  C_b &  0 \ea \, ,\]
and $G_g(\lambda)$ has only poles in $\mathds{C}_g$, while $G_b(\lambda)$ has only poles in $\mathds{C}_b$. The spectral separation in (\ref{desc-specdec}) is automatically provided by the Weierstrass canonical form of the pole pencil $A-\lambda E$, where the diagonal Jordan blocks are suitably permuted to correspond to the desired eigenvalue splitting.
This approach automatically leads to partial fraction expansions of $G_g(\lambda)$ and $G_b(\lambda)$.
A numerically reliable approach to compute spectral separations as in (\ref{desc-specdec}) has been proposed in \citep{Kags89}.

\subsection*{Coprime factorizations}

Consider a disjunct partition (\ref{Cgoodbadds}) of the complex plane $\mathds{C}$,
where both $\mathds{C}_g$ and $\mathds{C}_b$ are symmetrically located with respect to the real axis, and  such that $\mathds{C}_g$ has at least one point on the real axis. Any rational matrix $G(\lambda)$ can be expressed in a left fractional form
\be\label{lf} G(\lambda) = M^{-1}(\lambda)N(\lambda) \, , \ee
or in a right fractional form
\be\label{rf} G(\lambda) = N(\lambda)M^{-1}(\lambda) \, ,\ee
where both the denominator factor $M(\lambda)$ and the numerator factor $N(\lambda)$ have only poles in $\mathds{C}_g$. These fractional factorizations over a ``good'' domain of poles $\mathds{C}_g$ are important in various observer, fault detection filter, or controller synthesis methods, because they allow to achieve the placement of all poles of a TFM $G(\lambda)$ in the domain $\mathds{C}_g$ simply, by a premultiplication or postmultiplication of $G(\lambda)$ with a suitable $M(\lambda)$.

Of special interest are the so-called coprime factorizations, where the factors satisfy additional coprimeness conditions.  A fractional representation of the form (\ref{lf}) is a \emph{left coprime factorization} (LCF) of $G(\lambda)$ with respect to $\mathds{C}_g$, if there exist $U(\lambda)$ and $V(\lambda)$ with poles only in $\mathds{C}_g$ which satisfy the \emph{Bezout identity}
\[ M(\lambda)U(\lambda)+N(\lambda)V(\lambda) = I \, . \]
A fractional representation of the form (\ref{rf}) is a \emph{right coprime factorization} (RCF) of $G(\lambda)$ with respect to $\mathds{C}_g$, if there exist $U(\lambda)$ and $V(\lambda)$ with poles only in $\mathds{C}_g$ which satisfy
\[ U(\lambda)M(\lambda)+V(\lambda)N(\lambda) = I \, .\]

For the computation of a right coprime factorization of $G(\lambda)$ with a minimal descriptor realization $(A-\lambda E,B,C,D)$ it is sufficient to determine a state-feedback matrix $F$ such that all finite eigenvalues in $\Lambda(A+BF-\lambda E)$ belong to $\mathds{C}_g$ and all infinite eigenvalues in $\Lambda(A+BF-\lambda E)$ are simple. The descriptor realizations of the factors are given by
\[ \ba{c} N(\lambda) \\ M(\lambda) \ea = \ba{c|c} A+BF-\lambda E &B  \\ \hline C+DF & D \\  F & I_m \ea \, .\]
Similarly, to determine a left coprime factorization, it is sufficient to determine an output-injection matrix $K$ such that all finite eigenvalues in $\Lambda(A+KC-\lambda E)$ belong to $\mathds{C}_g$ and all infinite eigenvalues in $\Lambda(A+KC-\lambda E)$ are simple. The descriptor realizations of the factors are given by
\[ [\, N(\lambda) \; M(\lambda) \,] = {\arraycolsep=1mm\ba{c|cc} A+KC-\lambda E &B+KD & K  \\ \hline C & D  & I_p \ea }\, .\]

An important class of coprime factorizations is the class of coprime factorizations with minimum-degree denominators. The McMillan degree of  $G(\lambda)$ satisfies $\delta\left(G(\lambda)\right) = n_g + n_b$,
where $n_g$ and $n_b$ are the number of poles of $G(\lambda)$ in $\mathds{C}_g$ and $\mathds{C}_b$, respectively.
The denominator factor $M(\lambda)$ has the minimum-degree property if $\delta\left(M(\lambda)\right) =  n_b$. Special classes of coprime factorizations, as the coprime factorizations with inner denominators or the normalized coprime factorizations, have important applications in solving various analysis and synthesis problems. For the computation of coprime factorizations
with minimum degree denominators, descriptor system representation based methods  have been developed, which rely on iterative pole dislocation techniques \citep{Varg98a,Varg17d}.

\subsection*{Full rank compressions} \label{rccomp}

Row compressions of a $p\times m$ rational matrix $G(\lambda)$ of normal rank $r < p$ to a full row rank matrix can be determined by pre-multiplying $G(\lambda)$ with an invertible rational matrix $U(\lambda)$ to obtain
\[ U(\lambda)G(\lambda) = \ba{c} R(\lambda) \\ 0 \ea , \]
where $R(\lambda)$ has full row rank $r$. Of particular importance for solving model-matching problems is the case when $U(\lambda)$ has the form $U(\lambda) = Q^\sim(\lambda)$, where $Q(\lambda)$ is a square inner matrix, that is, $Q(\lambda)$ is stable and $Q^\sim(\lambda)Q(\lambda) = I$. If we partition $Q(\lambda)$ as $Q(\lambda) = [\, Q_{1}(\lambda) \; Q_{2}(\lambda) \,]$, with $Q_{1}(\lambda)$ having $r$ columns, then we have
\be\label{qrfac} G(\lambda) = Q(\lambda) \ba{c} R(\lambda) \\ 0 \ea = Q_{1}(\lambda)R(\lambda) .\ee
The full column rank  matrix $Q_1(\lambda)$ is an inner basis of the image space of $G(\lambda)$, while $Q_2(\lambda)$ is called its inner orthogonal complement.
We call (\ref{qrfac}) the \emph{inner--full-row-rank factorization} of $G(\lambda)$ and it can be interpreted as the generalization of the orthogonal rank-revealing QR factorization of a constant matrix.

The column compression of $G(\lambda)$ to a full column rank matrix can be obtained in a similar way, by post-multiplying $G(\lambda)$ with $Q^\sim(\lambda)$, where $Q(\lambda)$ is a square inner matrix. With $Q(\lambda)$ partitioned as $Q(\lambda) = \left[\begin{smallmatrix} Q_1(\lambda) \\ Q_2(\lambda) \end{smallmatrix}\right]$, with $Q_1(\lambda)$ having $r$ rows, then we can write
\be\label{rqfact}
G(\lambda) = [\, R(\lambda) \; 0 \,] Q(\lambda) = R(\lambda) Q_1(\lambda) .\ee
The full row rank matrix $Q_1(\lambda)$ is co-inner (i.e., $Q_1(\lambda)Q_1^\sim(\lambda) = I$) and is a basis of the co-image space of $G(\lambda)$. The factorization (\ref{rqfact}) is called the \emph{full-column-rank--co-inner factorization} and can be seen as a generalization of the orthogonal rank-revealing RQ factorization of a constant matrix.

The primary role of the inner matrix $Q(\lambda)$ is to achieve the row or column compression of $G(\lambda)$ to a full rank matrix. If $G(\lambda)$ has no zeros on the boundary of the stability domain $\mathds{C}_s$, then it is possible to achieve simultaneously that all zeros of $R(\lambda)$ result in the stable region $\mathds{C}_s$. Additionally, if $G(\lambda)$ is stable, then $R(\lambda)$ results stable too, and thus,  minimum-phase. In this case, the factorization (\ref{qrfac})
is called the \emph{inner-outer factorization} of $G(\lambda)$, with $R(\lambda)$ \emph{outer} (i.e., minimum-phase and full row rank), and the factorization (\ref{rqfact})
is called the \emph{co-outer--co-inner factorization} of $G(\lambda)$, with $R(\lambda)$ \emph{co-oute}r (i.e., minimum-phase and full column rank).
The inner-outer and co-outer--co-inner factorizations are instrumental in solving approximate controller and fault detection filter synthesis problems, which involve the minimization of $\mathcal{H}_\infty$-norm or $\mathcal{H}_2$-norm performance criteria. General methods to determine inner-outer factorizations are based on the computation of special Kronecker-like forms of the system matrix pencil \citep{Oara00,Oara05}.

\subsection*{Linear rational matrix equations}
The solution of model-matching problems encountered in the synthesis of controllers or filters involves  the solution of  the linear rational matrix equation
\be\label{lineq_EMMright} G(\lambda)X(\lambda) = F(\lambda) \, , \ee
with  $G(\lambda)$ a $p\times m$  rational matrix and $F(\lambda)$ a $p\times q$  rational matrix. This equation has a solution provided the compatibility condition
\be\label{compatibility} \rank G(\lambda) = \rank [\, G(\lambda)\; F(\lambda) \,] \ee
is fulfilled. Assume $G(\lambda)$ and $F(\lambda)$ have descriptor realizations of the form
\[ G(\lambda) = {\arraycolsep=1mm\ba{c|c}A-\lambda E & B_G \\ \hline C & D_G \ea, \;\; F(\lambda) = \ba{c|c}A-\lambda E & B_F \\ \hline C & D_F \ea} ,\]
which share the  system pair $(A-\lambda E,C)$.
It is easy to observe that any solution $X(\lambda)$ of (\ref{lineq_EMMright}) is also part of the solution $Y(\lambda) = \left[\begin{smallmatrix} W(\lambda) \\ X(\lambda) \end{smallmatrix}\right]$ of the linear (polynomial) equation
\be\label{lineq_pencil_right} S_G(\lambda) Y(\lambda) = \ba{c} B_F \\ D_F \ea \, , \ee
where $S_G(\lambda)$ is the associated system matrix pencil
\be\label{SysG} S_G(\lambda) = \ba{cc}A-\lambda E & B_G \\ C & D_G \ea . \ee
Therefore, an alternative to solving (\ref{lineq_EMMright}),  is to solve (\ref{lineq_pencil_right}) for $Y(\lambda)$ instead and compute $X(\lambda)$ as
\be\label{X_extracted_right}  X(\lambda) = [\;  0 \;\; I_p \;] Y(\lambda) \, .\ee
The compatibility condition (\ref{compatibility}) becomes
\[ {\arraycolsep=1mm\rank \ba{cc}A-\lambda E & B_G \\ C & D_G \ea =
\rank \ba{ccc}A-\lambda E & B_G & B_F \\ C & D_G & D_F \ea }.\]

If $G(\lambda)$ is invertible, a descriptor system realization of $X(\lambda)$ can be explicitly obtained as
\be\label{QinvF} X(\lambda)  = \ba{cc|c}  A-\lambda E & B_G & B_F\\ C & D_G & D_F\\ \hline 0 & -I_p & 0 \ea \, .\ee

The general solution of  (\ref{lineq_EMMright}) can be expressed as
\[ X(\lambda) = X_0(\lambda) + X_{r}(\lambda)Y(\lambda) , \]
where $X_0(\lambda)$ is any particular solution of (\ref{lineq_EMMright}),
$X_{r}(\lambda)$ is a rational basis matrix for the right nullspace
of $G(\lambda)$, and $Y(\lambda)$ is an arbitrary rational matrix with suitable dimensions. General methods to determine both $X_0(\lambda)$  and
$X_{r}(\lambda)$ can be devised by using the Kronecker canonical form of the associated system matrix pencil $S_G(\lambda)$ in (\ref{SysG}). It is also possible to choose $Y(\lambda)$ to obtain special solutions, as, for example, with least McMillan degree.  A numerically sound computational approach to determine least McMillan degree solutions is based on the reduction of the system matrix pencil $S_G(\lambda)$ to a Kronecker-like form and has been proposed in \citep{Varg04b}.

\subsection*{Approximate model matching}\label{appsec:AMMP}

We consider the following standard formulation of the approximate \emph{model-matching problem} (MMP): determine for a given stable $G(\lambda)$ and a stable $F(\lambda)$, a stable rational matrix $X(\lambda)$ such that
\[ \| F(\lambda) - G(\lambda)X(\lambda)\| = \min , \]
where either the $\mathcal{L}_2$-norm or $\mathcal{L}_\infty$-norm of the approximation error
$\mathcal{E}(\lambda) := F(\lambda)-G(\lambda)X(\lambda)$ are used.  The corresponding problems are called  $\mathcal{L}_2$-MMP and $\mathcal{L}_\infty$-MMP, respectively.

In the absence of general necessary and sufficient conditions for the existence of an optimal solution of the MMPs, an often employed  sufficient condition is to assume that  $G(\lambda)$  has no zeros on the boundary of  $\mathds{C}_s$. Furthermore, in the case of the $\mathcal{L}_2$-norm and for a continuous-time system, it is assumed that $F(s)$ is strictly proper.  These  conditions are clearly not necessary (e.g., if an exact solution exists).

The inner-outer factorization (\ref{qrfac}) of $G(\lambda)$, with $R(\lambda)$ outer,   can be employed to reduce the MMPs to simpler ones, the so-called  \emph{least distance problems} (LDPs).
The factorization (\ref{qrfac}) allows to express the error norm as
\be\label{errnorm}
\|\mathcal{E}(\lambda)\| = \left\|\ba{c} \widetilde F_1(\lambda)-Y(\lambda) \\ \widetilde F_2(\lambda)\ea\right\| \, ,\ee
where $Y(\lambda) := R(\lambda)X(\lambda)$ and
\[ Q^{\sim}(\lambda)F(\lambda)  = {\ba{c} Q_{1}^{\sim}(\lambda)F(\lambda) \\  Q_{2}^{\sim}(\lambda)F(\lambda) \ea := \ba{c} \widetilde F_1(\lambda) \\ \widetilde F_2(\lambda) \ea \, .} \]
The terms $\widetilde F_1(\lambda)$ and $\widetilde F_2(\lambda)$ are generally unstable, and may even be improper in the discrete-time case (i.e.,  if $Q(z)$ has poles in the origin).

The problem of computing a stable solution $X(\lambda)$ which minimizes the error norm $\|\mathcal{E}(\lambda)\|$ has been thus reduced to a LDP to compute the stable solution $Y(\lambda)$ which minimizes the norm in (\ref{errnorm}). The solution of the original MMP is given by
\[ X(\lambda) = R^{\dag}(\lambda)Y(\lambda)\, , \]
where $R^{\dag}(\lambda)$ is a stable right inverse of $R(\lambda)$ (i.e., $R(\lambda)R^{\dag}(\lambda) = I$).

The solution of the LDP in the case of $\mathcal{L}_2$-norm is straightforward. Let $L_s(\lambda)$ be the stable part and let $L_u(\lambda)$ be the
unstable part in the additive decomposition
\be\label{F2stabUnstab} \widetilde  F_1 (\lambda) = L_s(\lambda) + L_u(\lambda) \, ,\ee
where, in the continuous-time case, we take the unstable projection $L_u(\lambda)$  strictly proper.
The solution of the LDP  is
\[ Y(\lambda) = L_{s}(\lambda)  \]
and the achieved minimum  error norm of ${\mathcal{E}}(\lambda)$ is
\[ \| {\mathcal{E}}(\lambda)\|_2 =  \big\|\big[\,L_{u}(\lambda) \;\; \widetilde  F_2(\lambda)\,\big]\big\|_2
 \, .
\]

The solution of the LDP in the case of $\mathcal{L}_\infty$-norm is more complicated, and follows the approach described in \citep{Fran87} for continuous-time systems. The solution procedure  involves the solution of a Nehari problem and, if $\widetilde F_2(\lambda)$ is present (i.e., $G(\lambda)$ has no full row rank), the repeated solution of a special spectral factorization problem in a so-called $\gamma$-iteration approximation process. Details can be found in \citep{Fran87} for the continuous-time case, or in \citep[Chapter 9 and 10]{Varg17}.\\[-10mm]

\section{Software Tools}

Several basic requirements are desirable when implementing robust software tools for  numerical computations:
\begin{itemize}
\item employing exclusively numerically stable or numerically reliable algorithms;\\[-5mm]
\item ensuring high computational efficiency;\\[-5mm]
\item enforcing robustness against numerical exceptions (overflows, underflows) and poorly scaled data;\\[-5mm]
\item ensuring ease of use, high portability and high reusability.
\end{itemize}

The above requirements have been  enforced in the development of high-performance linear algebra software libraries, such as BLAS \citep{Dong90}, a collection of basic linear algebra subroutines, and LAPACK \citep{Ande99}, a comprehensive linear algebra package based on BLAS. These requirements have been also adopted to implement SLICOT \citep{Benn99,Huff04}, a subroutine library for control theory, based primarily on BLAS and LAPACK. The general-purpose library LAPACK contains over 1300 subroutines and covers most of the basic linear algebra computations for solving systems of linear equations and eigenvalue problems.  The release 4.5 of the specialized library SLICOT\footnote{\textbf{\url{http://www.slicot.org/}}} is a free software distributed under the GNU General Public License (GPL). The substantially enriched current release 5.7 is freely distributed under a BSD 3-Clause License via GitHub\footnote{\textbf{\url{https://github.com/SLICOT/SLICOT-Reference/}}} and contains over 500 subroutines. SLICOT covers the basic computational problems for the analysis and design of linear control systems, such as linear system analysis and synthesis, filtering, identification, solution of matrix equations, model reduction, and system transformations. Of special interest is the comprehensive collection of routines for handling descriptor systems and for solving generalized linear matrix equations, as well as, the routines for computing various Kronecker-like forms.
The subroutine libraries BLAS, LAPACK and SLICOT have been originally implemented in the general-purpose language Fortran 77 and, therefore, provide a high level of reusability, which allows their easy incorporation in user-friendly software environments, for example, MATLAB. In the case of MATLAB, selected LAPACK routines underlie the linear algebra functionalities, while  the incorporation of selected SLICOT routines was possible via suitable gateways, as the provided $mex$-function interface.

The Control System Toolbox (CST) of MATLAB supports both descriptor system state-space models and input-output representations with improper rational  TFMs, and provides a rich functionality covering the basic system operations and couplings, model conversions, as well as some advanced functionality such as pole and zero computations, minimal realizations, and the solution of generalized Lyapunov and Riccati equations. However, most of the functions of the CST can only handle descriptor systems with proper TFMs and  important functionality is currently lacking  for handling the more general descriptor systems with improper TFMs, notably for determining the complete pole and zero structures or for computing minimal order realizations, to mention only a few limitations.

To facilitate the implementation of the synthesis procedures of fault detection and isolation filters described in the book \citep{Varg17}, a new collection of freely available $m$-files, called the {Descriptor System Tools} (DSTOOLS), has been implemented for MATLAB.  DSTOOLS is primarily intended to provide an extended functionality for both MATLAB (e.g., with matrix pencil manipulation methods for the computation of Kronecker-like forms), and for the CST  by providing functions for  minimal realization of descriptor systems, computation of pole and zero structures, computation of nullspace and range space bases, additive decompositions, several factorizations of rational matrices (e.g., coprime, full rank, inner-outer), evaluation of $\nu$-gap distance, exact and approximate solution of linear rational matrix equations, eigenvalue assignment and stabilization via state feedback, etc.
The approach used to develop DSTOOLS exploits MATLAB's matrix and object manipulation features by means of a flexible and functionally rich collection of \emph{m}-files, intended for noncritical computations, while simultaneously enforcing highly efficient and numerically sound computations via \emph{mex}-functions (calling Fortran routines from SLICOT), to solve critical
numerical problems requiring the use of structure-exploiting algorithms.  An important
aspect of implementing DSTOOLS was to ensure that standard
systems are fully supported, using specific
algorithms. In the same vein, all algorithms are available
for both continuous- and discrete-time systems.

A precursor of DSTOOLS was the {Descriptor Systems} Toolbox for MATLAB, a  proprietary software of the German Aerospace Center (DLR), developed between 1996 and 2006 (for the status of this toolbox around 2000 see \citep{Varg00}). Some descriptor system functionality covering the basic manipulation of rational
matrices is also available in the free, open-source software Scilab\footnote{\textbf{\url{http:scilab.org}}} and Octave\footnote{\textbf{\url{http://www.gnu.org/software/octave/}}}. \\[-3mm]

A notable recent development is the free software package \textbf{\texttt{DescriptorSystems}}\footnote{\textbf{\url{https://github.com/andreasvarga/DescriptorSystems.jl}}},  which implements the complete functionality of  DSTOOLS in the Julia language. Julia is a powerful and flexible dynamic language, suitable for scientific and numerical computing, with performance comparable to traditional statically-typed languages such as Fortran or C. As a programming  language, Julia features optional typing, multiple dispatch, and good performance, achieved using type inference and just-in-time compilation \citep{Bezanson2017}. The  underlying Julia packages \textbf{\texttt{MatrixEquations}}\footnote{\textbf{\url{https://github.com/andreasvarga/MatrixEquations.jl}}}, for solving  various control related matrix equations (Lyapunov, Sylvester, Riccati), and \textbf{\texttt{MatrixPencils}}\footnote{\textbf{\url{https://github.com/andreasvarga/MatrixPencils.jl}}}, for manipulation of matrix pencils, 
provide the required basic computational functionality for the implementation of \textbf{\texttt{DescriptorSystems}}  (e.g., such as provided by SLICOT for DSTOOLS). A new feature of this package is the full support for models with both real and complex data \textit{(Note:} DSTOOLS supports only models with real data).

\section{Recommended Reading}

The theoretical aspects of descriptor systems are discussed in the two textbooks \citep{Dai89,Duan10}. For a thorough treatment of rational matrices in a system theoretical context two authoritative  references are \citep{Kail80} and \citep{Vidy11}. Most of the concepts and techniques presented in this article are also discussed in depth in \citep{Zhou96} for standard systems. The book \cite{Varg17} illustrates the use of descriptor system techniques to solve the synthesis problems of fault detection and isolation filters in the most general setting. Chapters 9 and 10 of this book describe in details the presented descriptor system techniques for the manipulation of rational matrices and also give  details on available numerically reliable algorithms. These algorithms form the basis of the implementation of the functions available in the DSTOOLS and  \textbf{\texttt{DescriptorSystems}} collections.
A comprehensive documentation of DSTOOLS is available in arXiv \citep{Varg18}.
A shorter version of this article appeared in the  Encyclopedia of Systems and Control \citep{Varg19d}.

\newpage

\end{document}